\documentclass[smallextended]{svjour3}
\usepackage{amsfonts}
\usepackage{latexsym}
\usepackage{amsmath}
\usepackage{amssymb}
\usepackage{amscd}
\usepackage{epsfig}
\usepackage{multirow}
\usepackage{graphicx}%, subfigure}
\addtocounter{MaxMatrixCols}{4}

\newcommand{\undertilde}[1]{\ensuremath{\mathord{\vtop{\ialign{##\crcr
   $\hfil\displaystyle{#1}\hfil$\crcr\noalign{\kern1.5pt\nointerlineskip}
   $\hfil\tilde{}\hfil$\crcr\noalign{\kern1.5pt}}}}}}
\begin{document}

\title{Using a Hamiltonian cycle problem algorithm to assist in solving difficult instances of Traveling Salesman Problem}
\author{V.~Ejov, M.~Haythorpe, S.~Rossomakhine}

\institute{Vladimir Ejov (ORCID: 0000-0002-5582-2859)
\at Flinders University, 1284 South Road, Tonsley, Australia 5042\\
\email{vladimir.ejov@flinders.edu.au} \and Michael Haythorpe (ORCID: 0000-0001-8143-6583) (Corresponding author)
\at Flinders University, 1284 South Road, Tonsley, Australia 5042, Ph: +61 8 8201 2375, Fax: +61 8 8201 2904\\
\email{michael.haythorpe@flinders.edu.au} \and  Serguei Rossomakhine
\at Flinders University, 1284 South Road, Tonsley, Australia 5042\\
\email{serguei.rossomakhine@flinders.edu.au}} \maketitle
{\abstract
We describe a hybrid procedure for solving the traveling salesman problem (TSP) to provable optimality. We first sparsify the instance, and then use a hybrid algorithm that combines a branch-and-cut TSP solver with a Hamiltonian cycle problem solver. We demonstrate that this procedure enables us to solve difficult instances to optimality, including one which had remained unsolved since its construction in 2002.

%\vspace*{-0.5cm}
\section{Introduction}\label{sec-Introduction}
%\vspace*{-0.3cm}

One of the most important problems in operations research is the {\em traveling salesman problem} (TSP), defined as follows. Given a set of cities $V$, and distances between each pair of cities, find a path that visits each city exactly once and returns to the starting city, such that the total distance travelled is minimised. A path which visits each city and then returns to the starting city is called a {\em tour}, and tours with the smallest distance are called {\em optimal tours}. A common way of representing instances of TSP is to simply provide a weighted graph, where the weight on edge $(i,j)$ is equal to the distance between cities $i$ and $j$. Since TSP was first posed in the 1930's, a vast amount of research has been carried out into how best to solve TSP, and despite TSP being an NP-hard problem, modern algorithms such as Concorde \cite{Concorde,ConcordeBook} and LKH \cite{LKH} are often capable of obtaining optimal tours in instances with tens of thousands of cities, although proving that optimality has been obtained is typically much more difficult.

Naturally, some instances are recognised as more difficult than others. It is important to note that \lq\lq difficulty" can have two different meanings in the context of TSP. An instance might be difficult if it is difficult to discover an optimal tour, or alternatively, it might be relatively easy to find the optimal tour but still difficult to confirm that it is, indeed, optimal. The majority of TSP heuristics make no attempt to confirm optimality of their discovered tours. Arguably, the best algorithm which does attempt to establish optimality is Concorde. The approach taken by Concorde is to find both lower and upper bounds for the length of the optimal tour, and then iteratively tighten these until they coincide. The lower bounds are produced by a branch-and-cut procedure, while the upper bound at any given stage is simply the length of the best tour discovered to that point by {\em linkern}, an implementation of the Lin-Kernighan heuristic \cite{LK}. As linkern is able to improve the upper bound, any branches of the branching tree with values larger than the new upper bound can be pruned. Of course, this tree grows in size exponentially, and so the running time of Concorde can grow exponentially as well.

One source of difficult instances of TSP arose from the consideration of the construction of computer chips with thousands of transistors, via a process known as very-large-scale integration (VLSI). Each of these instances, which can be found at \cite{cook}, takes the form of a chip containing transistors at various locations, and the distances between pairs of transistors are simply the Euclidean distances. Since their construction in 2002, many of these instances have been solved to optimality, but some remain unsolved. The smallest of these, called {\em dea2382}\footnote{It should be noted that since we reported the successful solving to optimality of dea2382, the VLSI Data Sets page \cite{cook} has been updated to indicate recently solved instances, including dea2382.}, has only 2382 vertices.

A related problem to TSP is the Hamiltonian Cycle Problem (HCP), which is the problem of determining whether {\em any} tours exist in a given graph. In this manuscript, we consider a hybrid algorithm produced by augmenting Concorde with an algorithm designed to solve HCP. We demonstrate the strength of this algorithm by solving many of the instances from \cite{cook} to optimality, including the previously unsolved instance {\em dea2382}.

\section{Sparsification Algorithms}

An valuable technique for solving large instances of TSP is to use a {\em sparsification algorithm}. In the context of weighted graphs, we say that a sparsification algorithm seeks to remove any edges which cannot be included in any optimal tour. For general TSP this is a difficult task; however, when the distances are Euclidean, there are various good sparsification algorithms available, such as the recent algorithm by Hougardy and Schroeder \cite{hougardy}.

If relatively few edges remain after a sparsification algorithm is applied, we say that the resulting instance is {\em sparse}. Removing these edges from consideration can be valuable for difficult instances. For instance, Hougardy and Schroeder claim that their sparsification enables the notoriously difficult instance d2103 to be solved eleven times faster in Concorde. However, in practice for instances of moderate difficulty, sparsification does not typically result in an improvement in solving time for Concorde, and in many cases is actually detrimental to the solving time. In order to run efficiently, Concorde relies on {\em linkern} obtaining reasonable upper bounds, so that its branching tree can be pruned. If the graph is sparse, it is often difficult for {\em linkern} to obtain any tour at all. To avoid this issue, Concorde will reintroduce any missing edges with a very large distance, so that the instance is complete. Then, {\em linkern} will certainly be successful at discovering a tour, but it is very likely that this tour will include at least one of these large-weight edges, and so the upper bound will also be large. Without a reasonable upper bound, Concorde's branching tree is permitted to grow to enormous size. When solving to optimality it is necessary to traverse the entire tree, and hence it is vital that this huge growth be avoided. It is therefore highly desirable to find {\em any} tour, even if it is sub-optimal, which uses only edges from the underlying sparse graph as early as possible to allow pruning of the tree to begin while its size is still manageable.

The problem of finding any tour in a sparse graph is equivalent to the {\em Hamiltonian cycle problem} (HCP) which simply asks, given a graph, whether any tours exist in the graph. By taking the sparsified instance of TSP, and ignoring the weights, we obtain an instance of HCP. If we are able to rapidly solve that instance of HCP, we can provide a tour to Concorde which will have a reasonable upper bound to prevent the branching tree from growing to unmanageable levels. Also, this tour can be used as a launching point for {\em linkern} which can then try to further improve the tour.

Although there are many HCP algorithms available (i.e. see \cite{chalaturnyk,eppstein,ali,frieze,gazettepaper,gondzio}), arguably the best HCP heuristic for difficult sparse instances is {\em Snakes and Ladders Heuristic} (SLH) \cite{SLH}. By augmenting Concorde with SLH, we are able to take advantage of SLH's ability to find tours in sparse graphs to restrict the growth of the branching tree. Once such a tour is found, {\em linkern} can then attempt to improve it. Of course, at this point, SLH can be run again, perhaps finding new tours which may, by chance, be better than the best known tour to that point. We refer to this augmented version of Concorde as {\em Concorde-SLH}. It should be noted that Concorde-SLH is only useful if the instance has been sparsified first.

\section{Results}

We now demonstrate the merit of the approach described above by considering fifteen instances from \cite{cook}. These are the fifteen instances containing between 1500 and 2500 vertices, and includes the instance {\em dea2382} which has not previously been solved to optimality. After sparsifying the instances using the sparsification algorithm by Hougardy and Schroeder, and submitting these sparsified instances to Concorde-SLH, we were successful in solving each of the instances to optimality. To demonstrate the improvement obtained by using Concorde-SLH, in Table \ref{tab-results} we compare the times taken to those using the standard implementation of Concorde for both the original (non-sparsified) instances, and the sparsified instances. Any instances which took longer than 100 hours to solve were terminated early. Of the fifteen instances, {\em dea2382} took the longest, with Concorde-SLH taking just over one day to terminate. Concorde and Concorde-SLH were run on a cluster computer network where the individual machines each contained an AMD Opteron 6282 SE @ 1.4Ghz (4 cores, 8 threads) and 16GB DDR3 RAM. The version of Concorde used was the binary executable version 03.12.19.

\begin{table}[h!]\begin{center}\begin{tabular}{|r|c|c|c|c|c|c|c|}
%\hline {\bf Instance} & {\bf Concorde Time} & {\bf Concorde-SLH Time} & {\bf Optimal Tour Length}\\
%\hline  & {\bf Concorde} & {\bf Concorde-SLH} & & {\bf Concorde} & {\bf Concorde-SLH}\\
\hline & {\bf Concorde} & {\bf Concorde} & {\bf Conc-SLH} & & {\bf Concorde} & {\bf Concorde} & {\bf Conc-SLH} \\
{\bf Instance} & {\bf (Original)} & {\bf (Sparse)} & {\bf (Sparse)} & {\bf Instance} & {\bf (Original)} & {\bf (Sparse)} & {\bf (Sparse)} \\
\hline {\bf rbv1583} & 00:06:08 & 00:17:02 & 00:04:05 & {\bf bva2144} & 00:47:57 & 00:43:38 & 00:08:26 \\
\hline {\bf rby1599} & 38:51:45 & Crash & 02:45:10 & {\bf xqc2175} & 87:28:16 & Crash & 04:50:34 \\
\hline {\bf fnb1615} & 00:19:36 & 00:39:48 & 00:08:04 & {\bf bck2217} & Timeout & Timeout & 03:33:15 \\
\hline {\bf djc1785} & 03:02:29 & 03:52:24 & 00:42:25 & {\bf xpr2308} & Timeout& Crash & 05:14:44 \\
\hline {\bf dcc1911} & 04:51:32 & 04:40:32 & 00:24:34 & {\bf ley2323} & 00:41:37 & 00:22:52 & 00:04:36 \\
\hline {\bf dkd1973} & 01:07:38 & 05:15:33 & 00:26:48 & {\bf dea2382} & Timeout & Timeout & 31:44:19 \\
\hline {\bf djb2036} & 45:31:27 & Timeout & 01:15:05 & {\bf rbw2481} & 00:15:04 & 00:29:17 & 00:06:11 \\
\hline {\bf dcb2086} & 00:29:30 & 01:54:27 & 00:09:15 & & & & \\
%\hline {\bf bva2144} & & &\\
%\hline {\bf xqc2175} & & &\\
%\hline {\bf bck2217} & & &\\
%\hline {\bf xpr2308} & & &\\
%\hline {\bf ley2323} & & &\\
%\hline {\bf dea2382} & Timeout & 76:44:19 & 8017\\
%\hline {\bf rbw2481} & & &\\
\hline\end{tabular}\caption{Time taken (hh:mm:ss) for both Concorde, and Concorde-SLH, to solve sparsified instances of TSP to optimality, and the lengths of the optimal tours. Timeout means the run took longer than 100 hours to complete. Crash means the algorithm terminated early due to an issue with the LP solver.}\label{tab-results}\end{center}\end{table}

As can be seen in Table \ref{tab-results}, sparsifying the instances did not lead to significant improvement for Concorde. Indeed, it often resulted in a slower computation time, or even a crash when the LP solver encountered a problem; this is a known issue that arises occasionally for difficult non-complete instances. However, the addition of SLH resulted in a significantly improved solving time in all tested sparsified instances. This include the successful solving to optimality of the previously unsolved instance {\em dea2382} in a little over one day of computation time.

\bibliographystyle{plain}   % (uses file "plain.bst")
%\bibliography{biblio}       % expects file "myrefs.bib"

\begin{thebibliography}{99}
\bibitem{Concorde} Applegate, D.L., Bixby, R.B., Chav\'{a}tal, V., and Cook, W.J. Concorde TSP Solver: http://www.tsp.gatech.edu/concorde/index.html.
\bibitem{ConcordeBook} Applegate, D.L., Bixby, R.B., Chav\'{a}tal, V., and Cook, W.J. {\em The Traveling Salesman Problem: A Computational Study}. Princeton University Press (2006).
\bibitem{SLH} Baniasadi, P., Ejov, V., Filar, J.A., Haythorpe, M., and Rossomakhine, S. Deterministic \lq\lq Snakes and Ladders" Heuristic for the Hamiltonian cycle problem, {\em Math. Program. Comput.}, 6(1):55-75, 2014.
\bibitem{chalaturnyk} Chalaturnyk, A. {\em A Fast Algorithm For Finding Hamiltonian Cycles.} Ph.D Thesis, University of Manitoba, 2008.
\bibitem{cook} Cook, W.J. VLSI Data Sets. http://www.math.uwaterloo.ca/tsp/vlsi/, 2013.
\bibitem{eppstein} Eppstein, E. The traveling salesman problem for cubic graphs, {\em J. Graph Algorithm. App.}, 11(1):61--81, 2007.
\bibitem{ali} Eshragh, A., Filar, J.A., and Haythorpe, M. A hybrid simulation-optimization algorithm for the Hamiltonian cycle problem, {\em Ann. Oper. Res.}, 189(1):103--125, 2011.
\bibitem{frieze} Frieze, A., and Haber, S. An almost linear time algorithm for finding Hamilton cycles in sparse random graphs with minimum degree at least three, {\em Random Struct. Algor.}, 47(1):73--98, 2015.
\bibitem{gazettepaper} Haythorpe, M. Finding Hamiltonian Cycles Using an Interior Point Method, {\em Austral. Math. Soc. Gaz.}, 37(3):170--179, 2010.
\bibitem{gondzio} Ejov, V., Filar, J.A., and Gondzio, J. An Interior Point Heuristic for the Hamiltonian Cycle Problem via Markov Decision Processes, {\em J. Global Opt.}, 29(3):315--334, 2004.
\bibitem{LKH} Helsgaun, K. An Effective Implementation of Lin-Kernighan Traveling Salesman Heuristic. {\em Eur. J. Oper. Res.} 126, 106--130 (2000).
\bibitem{hougardy} Hougardy, S., and Schroeder, R.T. Edge elimination in TSP instances. In: {\em International Workshop on Graph-Theoretic Concepts in Computer Science}. Springer International Publishing, 2014.
\bibitem{LK} Lin, S. and Kernighan, B.W. An Effective Heuristic Algorithm for The Traveling Salesman Problem. {\em Oper. Res.} 21, 496--516 (1973).

\end{thebibliography}

\end{document}